# Progress Report on a Proposed Theory for Software Development

Diana Kirk[1] and Stephen G. MacDonell[2]
[1]*Consultant, Auckland, New Zealand*
[2]*Software Engineering Laboratory (SERL)*
*Auckland University of Technology (AUT), Auckland, New Zealand*

**Abstract**

*There is growing acknowledgement within the software engineering community that a theory of software development is needed to integrate the myriad methodologies that are currently popular, some of which are based on opposing perspectives. We have been developing such a theory for a number of years. In this position paper, we overview our theory along with progress made thus far. We suggest that, once fully developed, this theory, or one similar to it, may be applied to support situated software development, by providing an overarching model within which software initiatives might be categorised and understood. Such understanding would inevitably lead to greater predictability with respect to outcomes.*

## 1. INTRODUCTION

The term *Software Engineering* was coined in 1968 at a conference whose aim was to discuss the need for the software development discipline to be more strongly based on theoretical and engineering principles (Naur and Randell, 1969). The *Waterfall* model, a then-popular model used in manufacturing, was adopted as the standard approach for developing computer software. As time progressed, it became apparent that a strict implementation of this model was not appropriate for software. A number of modifications, for example *Spiral* (Boehm, 1988), and alternative models, for example *XP* (Beck, 2000), have emerged. The authors of the various models have different viewpoints on what kind of activity software development actually *is*. Earlier models view soft- ware development as an engineering activity and focus on control. More recent models adopt the viewpoint of 'software-as-a-service' and focus on effective communications. However, regardless of the huge variation in approach, until recently, the accepted wisdom by all methodology architects was that, in order to be fully effective, their approach must be followed exactly, with nothing added and nothing missed (Cusumano et al., 2003).

We have long understood from experiences in industry that this 'wisdom' is not based on what actually happens in the field, and have advocated with others the need to more deeply understand the process of developing computer software in order to support industry in its need to select practices in a flexible way, according to objectives and context (Bajec et al., 2007; Fitzgerald, 1997; Hansson et al., 2009; Kirk and Tempero, 2004; Kirk and Tempero, 2005; Kirk, 2007). This viewpoint has now become the accepted one (Avison and Pries-Heje, 2008; MacCormack et al., 2012; Petersen and Wohlin, 2009; de Azevedo Santos et al., 2011; Turner et al., 2010). The traditional viewpoint — that methodologies and practices should be adopted and used as prescribed — has thus been superseded by one of acceptance that tailoring is both necessary and unavoidable.

If tailoring is of the essence, we clearly must strive to fully understand the nature of the relationships between objectives, process and context. Only then will we be in a position to advise industry on which practices might be most suitable or to predict project outcomes based on context and practices implemented. The sole route to this kind of general understanding is through theory-building (Gilmore, 1990). Without an understanding of the relationships between objectives, practices, context and outcomes, we can at best expose patterns based on data or observations. Such patterns represent *correlations* and correlations can not be used to predict in a general way. For consistent prediction, we must create frameworks based on causal relationships i.e. theoretical models. Indeed, Basili et al. remind us that, when carrying out controlled experiments, "... it is hard to know how to abstract important knowledge without a framework for relating the studies" (Basili et al., 1999).

The role of theory in software engineering (SE) has been investigated from a number of perspectives. Sjøberg et al. observed that there is very little focus on theories in software engineering and remind us of the key role played by theory-building if we wish to accumulate knowledge that may be used in a wide range of settings (Sjøberg et al., 2008). Hannay et al. conducted a review of the literature on experiments in SE and found that fewer than a third of studies applied theory to explain the cause-and-effect relationship(s) under investigation, and that a third of the theories applied were themselves proposed by the article authors (Hannay et al., 2007). Gregor examined the nature of theory in Information Systems (IS) and found multiple views on what constitutes a theory (Gregor, 2006). Stol and Fitzgerald argue that SE research does, in fact, exhibit traces of theory and suggest that the current focus on evidence based software engineering (EBSE) must be combined with a theory-focussed research approach to support explanation and prediction (Stol and Fitzgerald, 2013).

While there is general agreement within the research community that an increased focus on theory building would produce benefits, there remains uncertainty about how to proceed. Wand et al. remind us that "To employ conceptual modeling constructs effectively, their meanings have to be defined rigorously" and that achievement of this requires an articulation within the context of *ontology* (Wand et al., 1999). The authors apply an ontological framework developed by Bunge (Bunge, 1977) to analyse the relationship construct in entity-relationship modelling. A number of authors have used their approach to investigate various aspects of IS, for example, UML (Opdahl and Henderson-Sellers, 2002) and reference models (Fettke and Loos, 2003). A key aspect of the approach is the ability to confirm that a model is complete and without redundancy.

Our overall research objective is to apply an ontological approach in the development of a conceptual model to describe a *software initiative*. By *software initiative*, we mean the software-related processes implemented to achieve specified outcomes. We suggest that the existence of such a model would support the software industry in the selection of appropriate practices, according to an organisation's specific objectives and contexts. However, before we are in a position to formally conceptualise a software initiative, we must gain a better understanding of the constructs that will form the basis of our model. In this paper, we present our progress towards this deeper understanding. In section 2, we overview our progress thus far. In section 3 we discuss other efforts towards providing a theoretical foundation for software development and in section 4, we summarise the paper and discuss limitations and future work.

## 2. THEORY OVERVIEW

Our first observation is that we must scope our model to include *any* software initiative i.e. the scope is larger than a software *project*. Our rationale is the need to consider in a holistic way the entire *software- in-a-system (SIAS)* i.e. the software product or service as part of a larger whole. This 'whole' will vary in time as the software is first created and then used. The larger system during creation will include the development organisation, test environments (possibly standalone) and the client. After deployment, the software will become part of a system comprising any of hardware, software, humans and processes. The job is to create, deliver and sustain healthy software systems over a lifetime of operation. The reason behind our viewpoint of a need for greater holism has its roots in the uncertainty that results from the growing complexity of software systems. This complexity makes it impossible to categorise in a simple way the environment during development and makes it difficult to anticipate all future conditions under which the software will run i.e. we can no longer assume a stable and bounded operating environment. Conditions such as technology change and inappropriate use will probably affect the in-situ efficacy of the software. Uncertainty also characterises delivery mechanisms — in the past, a *project* developed a software product and then delivered this to a known client base. More recently, the web-based mechanism of 'deliver a little, solicit immediate client feedback, and deliver the next increment' has become popular (Stuckenberg and Heinzl, 2010), leaving the concept of a 'project- to-develop-then-deliver' no longer tenable.

As early as 1987, Basili and Rombach believed that "Sound tailoring requires the ability to characterise ... goals ..., the environment ..., and the effect of methods and tools on achieving these goals in a particular environment" (Basili and Rombach, 1987). Our efforts thus far have focused on clarifying what this might mean when considering flexibility in soft- ware initiatives. We thus begin with a consideration of *objectives*, *process* and *context*. In the following sections, we overview our efforts and comment on understandings achieved.

### 2.1 Objectives
We first observe the need to consider more than one objective for an initiative (Kirk and Tempero, 2005; Kirk et al., 2009). The reason is that a focus on a single outcome may lead to the identification of a local maximum and a possible sub-optimisation of the whole system (Kitchenham et al., 2002; Lakey, 2003; Lehman, 1997). We next observe that there are many possible objectives, including the common product-related ones of minimisation of cost and maximisation of quality, but also including the people-related ones such as 'increase developer subject-area knowledge', 'retain developers' or 'keep a specific customer happy'. For some of these, an associated numerical value will change throughout the initiative, for example, when spending increases throughout a project. For others, the goal is less definitive and a more fuzzy measure may be appropriate, for example, developer satisfaction levels may be described as 'Low', 'Medium' and 'High' (Kirk and MacDonell, 2009). A third observation is that most software initiatives are characterised by uncertainty (Atkinson et al., 2006; Perminova et al., 2007) and this means that values cannot be represented in a deterministic way. A probabilistic distribution is a more suitable measure for some kinds of objective (Connor, 2007; Kitchenham and Linkman, 1997; Rao et al., 2008). Our proposal is that a set of objectives may be modelled by a set of *values*, the type of value for each depending upon the nature of the objective. An implementation would involve representing an objective by a name, a description and a desired value.

Monitoring of progress, i.e. 'current state', will involve a consideration of the current value for each objective in the set. This takes into account the fact that state values are generally not 'empty' at commencement. For example, developers will have a certain 'satisfaction' level at commencement; in a product line situation, there is an existing code base that is characterised by a level of quality. We have as yet taken this investigation no further i.e. the structure of *objectives* is an open research question.

## 2.2 Process

We view a *process* as a set of *practices*. Each *Practice* has the effect of moving the initiative closer to (or further away from) its *Objectives*. We observe that, in addition to the accepted practices such as 'design inspection', our definition includes anything that has the desired effect. For example, in a small startup organisation, an informal, unplanned practice such as 'chat over lunch' may be crucial for supporting the developer's understanding of what is to be built and is thus an acceptable 'practice' in our model. According to our model, an effective practice is one that successfully moves the initiative in the right direction. A 'lean' process is one in which every practice is effective.

The space of all possible practices is huge, and includes practices for identifying the target audience for a proposed product and understanding what the product should do, practices for designing, implementing and delivering the product, and practices for supporting product use in the target environment(s). We clearly need to introduce some structure, but found that the standard reference models did not suit when attempting to elicit information from individuals in smaller, less formal organisations (Kirk and Tempero, 2012a). It was clear that, if we wanted to capture practices-as-implemented-in-the-real-world, some new perspective was required.

Our approach considers what organisations *need to achieve* at a high level when involved in a software initiative. Our top-level functional categories are

- Define what is to be made
- Make it
- Deliver it

We extended these categories to be what we believe are the main sub-categories for software. These are shown in table 1.

Table 1. Categories for practices.

| | |
|---|---|
| **Define** | Roadmap |
| | Scope |
| **Make** | Design |
| | Implement |
| | Integrate |
| **Deliver** | Release |
| | Support |

Because we have structured based on *function*, the categorisation will support any practice deemed to be relevant for meeting objectives. Informal meetings in the lunch room that help developers understand scope clearly fit into the 'scope' category. We hypothesise that the proposed categorisation addresses *all* software practices. A precondition that an objective be met is that each category must contain one or more effective practices. To illustrate, for a 'Quality' objective, including quality considerations during product design, implementation and integration will fail to achieve the objective if quality expectations are not included during scoping. The identification of gaps is straightforward.

The categorisation above does *not* imply any ordering of practices. Such ordering would exist at a higher level and might be used to describe strategies of iteration and incremental delivery. We also submit that the categorisation is 'paradigm-agnostic' — whether an initiative is run in a traditional or agile way, the basic functions of defining, making and delivering the product must be carried out. Of note is the fact that the sub-categories 'Roadmap' and 'Support' lie outside of a traditional development project and sub-categories 'Scope' and 'Release' relate to practices that span development organisation and client.

Our work on the *practices* aspect of the proposed theory is embryonic. Testing thus far is limited to a single, exploratory study in which we captured practices in three New Zealand software organisations (Kirk and Tempero, 2012a; Kirk and Tempero, 2012b). We did expose some interesting areas for further study. For example, all participating organisations reported a dependence on practices that involved having to actively search for information, possibly implying some inefficiency as individuals must spend time. However, the study was exploratory in nature and this is an open research area.

## 2.3 Context

This represents the most challenging aspect of any theory for software development. There have been many attempts to relate project outcomes to specific contextual factors, for example, (Avison and Pries-Heje, 2008; Clarke and O'Connor, 2012; Kruchten, 2013; Stuckenberg and Heinzl, 2010). Our main critique of existing approaches is that they remain factors based (Kirk and MacDonell, 2013). We suggest that such an approach is misguided because

- there are simply too many possible factors to take into consideration, and this number will increase as new paradigms for software are introduced.
- it is unlikely that any two projects will be exactly the same and so understanding key factors for some is unlikely to be of use in a general way.

We believe that it is crucial that we develop an operationalisation of *context* that will be relevant for all software initiatives i.e. takes into account the situated nature of a software product throughout its life-time. It seems clear that the required model must comprise a number of orthogonal dimensions in order that an initiative can be plotted as a point in the dimensional space. In order to remove the 'factors-based' aspect, it is necessary to also find suitable abstractions for each dimension, abstractions that support a straight-forward identification of value for a given initiative. For example, rather than define a number of factors such as 'developer experience', 'developer subject area knowledge', we must abstract in such a way as to render it irrelevant if we have missed a factor out (for example, 'developer commitment').

Our first efforts at modelling this space involved a consideration of the dimensions *Who*, *Where*, *What*, *When*, *How* and *Why* (Kirk and MacDonell, 2013). These dimensions have been applied by others to ensure orthogonality (Dybå et al., 2012; Zachman, 2009). Of course, the usefulness of the abstraction de- pends upon the choices about what these dimensions *mean*. Our assigned meanings are (Kirk and Mac- Donell, 2014b):

**Who** associates with peoples' *ability to perform*. Personal characteristics, culture and group structure are relevant, as these affect levels of under- standing and conceptual sharing.

**Where** associates with peoples' *availability*. The degrees of temporal and physical separation are relevant.

**What** associates with *product* characteristics. Standards expectations, product interfaces and achieved quality are relevant.

**When** associates with *product life cycle*. Examples are in-development, recently deployed, near retirement (MacDonell et al., 2008).

**How** associates with *engagement expectations*. Client and developer expectations for the mechanisms for product specification and delivery will affect which practices are most appropriate.

**Why** associates with *establishing objectives*.

We carried out some 'proof-of-concept' studies on this model, each involving categorising contextual factors from a small number of studies from the software engineering literature (Kirk and MacDonell, 2014b; Kirk and MacDonell, 2014a). These studies caused us to refine our understanding of context in the following ways. In the first instance, it became clear that the dimension *why* addresses *objectives* and is thus not part of a model for *context*. We also found that many contextual factors mentioned in the literature are vague or ambiguous, and so must be clarified prior to categorising. For example, when considering the commonly-mentioned factor 'Company size', we suggest that it is not size itself that affects practice selection and/or outcomes, but rather what this means in terms of culture and physical and temporal separation. We labelled such factors as 'Secondary'. Some factors, such as 'requirements uncertainty', may be the result of one of a number of possible scenarios. For example, perhaps the client is not clear about what the product exactly is; perhaps (s)he is simply weak on decision-making; perhaps (s)he is unable to state what is wanted because of client-internal processes i.e. (s)he is waiting for a decision from someone else.

We cannot know which practices will be effective until we understand which meaning is relevant. A practice of 'regular client meetings with prototypes' will not help if the client is waiting for someone else. We labelled this kind of factor as 'ambiguous'. Finally, we noticed that some factors were more 'high level' in that they could be more usefully viewed as affecting strategy. For example, 'lack of funds' would likely force some consideration about strategy, and the resulting decision may, in turn, affect objectives and/or context. It might be decided that the project should be abandoned, that some developers should be removed, or that a minimal product only should be delivered. We recognise these factors as *Strategic factors* and remove from our discussion of *context*.

Our current status is that, with our refined definition of *context*, we are well into a study to categorise contextual factors from the literature into our dimensional model. Thus far, we have met no obstacles. However, we have as yet included only literature from the *software engineering* domain and have constrained the study to the *development project*. Although we are optimistic, there clearly is much scope for research in this area, research that must be carried out before we can be confident in our proposed structure for *context*.

## 3. RELATED WORK

As far as we are aware, the only initiative to create a theory for software is the SEMAT initiative, launched in 2009 by Ivar Jacobson, Bertrand Meyer and Richard Soley (Jacobson et al., 2013a). The approach proposes a *SEMAT kernel* comprising three parts. The first is a means of measuring project progress and health, the second categorises the activities required to effect progress, and the third defines the competencies required to effect the activities (Jacobson and Seidewitz, 2014). There are seven top- level 'alphas' - *Requirements*, *Software System*, *Work*, *Team*, *Way of Working*, *Opportunity* and *Stakeholders*. It is claimed that these concepts support determination of a project's health and facilitate selection of a suitable set of practices.

Although the authors of the SEMAT approach state that the initiative promotes a "non-prescriptive, value-based" philosophy that encourages selection of practices according to context, we suggest that the approach is, in fact, prescriptive in intent. Each of the SEMAT elements has a number of states "which may be used to measure progress and health" (Jacobson et al., 2013b), and this implies that the health of every software project can be measured in a common way.

We do not believe this to be the case, because of the vast number of possible objectives and contexts. In addition, the SEMAT model appears to offer no guidance on *how* to choose activities. Other than mentioning competencies required for activities, there is no link between activities and objectives or context. How do we know if an activity is suitable when the developers are in different countries or when the team comprises multiple cultures? The authors state that gaps and overlaps in activities can be easily identified, but it is not clear how this can be achieved in an objective way. We also note that the theory relates to a software *project*, a scope we have identified as being too narrow. For the above reasons, we have issue with the claim that a general theory is being developed. We cannot see how the approach will help researchers better understand practice limitations. Weiringa, discussing the field of requirements engineering, reminds us of the dangers of confusing solution design with research (Wieringa, 2005). We suggest that SEMAT, at this point, represents a design initiative rather than a theory building exercise.

## 4. DISCUSSION

One characteristic of a successful theory is that opposing viewpoints can be more deeply understood and can be seen to be part of a larger picture. Our conceptualisation must address the traditional versus agile dichotomy. It is not difficult to see that, as a *practice* is defined simply as a transformation of current state, many different practices will exist that implement the same kind of transformation. For example, in relation to table 1, the practices 'Formally document requirements' and XP's 'Planning Game' both result in an in- creased understanding of what is to be built. However, if we consider the objectives 'Reliability' and 'Increase developer subject area knowledge', it is likely that the formal documentation approach will address the former (as non-functional requirements are an inherent part of formal requirements documents) while 'Planning Game' will not. The XP approach may require additional practices to address quality expectations.

Thus far, we have aimed to gain a better understanding of what appear to be the key constructs required i.e. *objectives*, *process* and *context*. From the perspective of the Bunge model, we suspect that *objectives* and *context* represent basic 'things'. However, a number of aspects are less clear, for example, the role of *process*. One possibility is that it may be represented as a basic 'thing'. It is more likely that the constituent *practices* and *context* mutually affect each other and this situation is represented as a 'mutual property' in the Bunge scheme. Our analysis above also exposed the need for a new idea, that of *Strategic factor* and we do not yet understand what this is from an ontological perspective. We do not under- stand how to represent the dimensions we propose for *context* — are these *properties* or *constituent components* of context? What are the repercussions of the decisions made? This is an exciting step on our journey, one which we now feel ready to address.

In this paper, we have presented an overview of the theoretical approach we have been pursuing for several years. Our work is in-embryo. Our contribution is that we are making an honest attempt to tackle an extremely difficult problem, and believe we have made pockets of progress in some areas. Our hope is that the efforts we have made thus far will be used as a starting point for other researchers. We submit that, if the community does not take the theory-building initiative seriously, we are doomed to endless cycles of 'new' process paradigms and architectures, each of which has some merit and many shortfalls.